\UseRawInputEncoding

\documentclass[aps,prl,twocolumn,amsmath,amssymb,nofootinbib,superscriptaddress,floatfix]{revtex4-1}

\usepackage[utf8]{inputenc}
\usepackage{gensymb}
\usepackage{amsmath}
\usepackage{amssymb}
\usepackage{amsthm}
\usepackage{mathtools}
\usepackage{mathdots}
\usepackage{amsfonts}
\usepackage{bbm}
\usepackage{xr}
\usepackage{bbold}
\usepackage{newpxtext}
\usepackage{epsfig,color,dsfont,upgreek,physics}
\usepackage{mathrsfs}
\usepackage{multirow}
\usepackage[makeroom]{cancel}
\usepackage{mathtools}

\usepackage{graphicx}
\usepackage{dcolumn} 
\usepackage{bm} 
\usepackage{float} 
\usepackage[driverfallback=dvipdfm]{hyperref} 
\usepackage{longtable}
\usepackage{ulem}   
\allowdisplaybreaks

\newcommand{\bs}[1]{{\boldsymbol{#1}}}

\begin{document}

 \title{Anyon bound states in \texorpdfstring{MoTe$_2$}{MoTe2} Fractional Chern Insulators
}

\author{Tianhong Lu}
\affiliation
{
Department  of  Physics,  Emory  University,  400 Dowman Drive, Atlanta,  GA  30322,  USA
}
\author{Luiz H. Santos}
\affiliation
{
Department  of  Physics,  Emory  University,  400 Dowman Drive, Atlanta,  GA  30322,  USA
}

\begin{abstract}
Binding anyons to optically active excitons offers a route to manipulate fractional excitations and reveal the internal spatial structure of anyons. We establish the formation of anyon bound states between a mobile dipolar exciton and an $e/3$ quasihole in a twisted bilayer MoTe$_2$ fractional Chern insulator. Using exact diagonalization of a continuum-based interacting model, we identify a regime of meV-scale binding and strong exciton-quasihole entanglement controlled by exciton mobility and the distance between the exciton and the FCI. Spatial correlations resolve a localized depletion core carrying approximately $e/3$ charge around the exciton, providing a real-space characterization of the bound state. Beyond its fractional charge, the quasihole exhibits a moiré-induced octupolar structure. Interlayer tunneling anisotropy strongly deforms this structure while leaving the enclosed charge nearly unchanged, linking the internal profile of the anyon to the underlying electronic quantum geometry. These findings establish conditions for robust exciton-anyon binding in moiré heterostructures and show how these composites provide access to properties of fractional excitations beyond their total charge.
\end{abstract}
\date{\today}

\maketitle

\noindent
\textit{Introduction--}
The realization of fractional Chern insulators (FCIs) in twisted bilayer MoTe$_2$ \cite{cai2023signatures,zeng2023thermodynamic,park2023observation,xu2023observation} 
and rhombohedral multilayer graphene \cite{lu2024fractional} has renewed interest in phases of matter that host anyons \cite{leinaas1977theory, wilczek1982quantum,cao2025fractional,lu-santos-2024fractional,lu2025electromagnetic,wang2024fractional,reddy2023fractional,yu2024fractional,li2021spontaneous}. These emergent excitations carry fractional charge and obey exchange statistics distinct from those of bosons and fermions, making them of interest for topological quantum information \cite{nayak2008non,stern2008anyons}. Harnessing them requires the ability to trap and manipulate individual anyons, yet their internal structure remains difficult to probe: unlike elementary particles, anyons are extended objects arising from a collective rearrangement of many electrons. How can anyons be pinned, and what internal properties do they possess in moiré FCI systems?

A promising route has recently emerged: anyonic quasiholes are predicted to bind to dipolar excitons in FQH systems \cite{mostaan2025anyon,wagner2025sensing} and in FCIs
\cite{lu2026exciton}, and signatures of closely related anyon--trion bound states have been observed experimentally in moir\'e FCIs \cite{li2026signatures}. These ideas connect to earlier and ongoing work on anyons bound to impurities \cite{grusdt2016interferometric,wagner2025sensing,wang2026anyon,liu2025characterization}.
These developments open the prospect of using optically induced excitonic complexes to spatially manipulate fractional quasiparticles and probe their internal structure. This prospect is especially timely for transition metal dichalcogenide (TMD) heterostructures, where FCIs and excitons naturally coexist and the populations of
fermionic and bosonic species can be controlled.

\begin{figure}[!t]
    \centering
    \includegraphics[width = \linewidth]{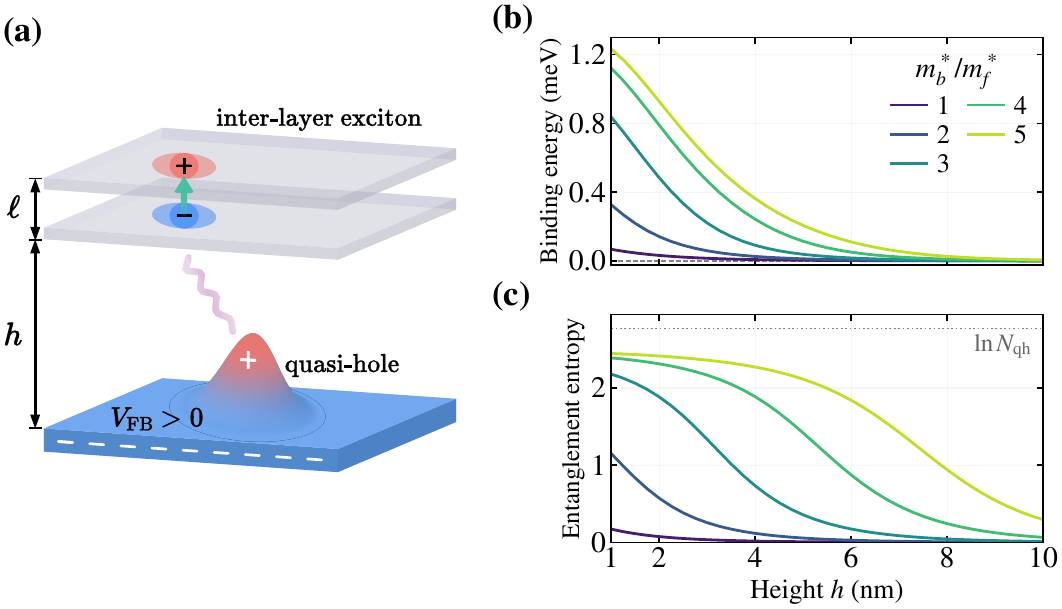}
\caption{Exciton-quasihole binding and entanglement. (a) Schematic of an interlayer exciton coupled to a fractional Chern insulator. The exciton’s constituent charges are separated by $\ell$, and its lower charge lies at distance $h$ above the electronic layer. The repulsive electron–exciton interaction $V_{\mathrm{FB}}>0$ effectively attracts a quasihole. (b) Binding energy versus $h$. (c) Fermi-Bose bipartite entanglement entropy $S_{\rm fb}=-\mathrm{Tr}(\rho_b\ln\rho_b)$ of the exciton-quasihole ground state. The dotted line marks $S_{\rm max}=\ln N_{\rm qh}$, where $N_{\rm qh}=16$ is the number of quasihole momentum sectors. 
Colors denote the boson-to-fermion effective-mass ratio $m_b^*/m_f^*$, as indicated in (b). Calculations use five fermions on a $3\times5$ lattice for the $\nu=1/3$ FCI and a $4\times4$ lattice containing one $e/3$ quasihole.}
\label{fig: Fig 1}
\end{figure}

In this Letter, we characterize the formation of exciton--anyon bound states in heterostructures in which twisted bilayer MoTe$_2$ (t-MoTe$_2$) is placed in proximity to a TMD heterobilayer hosting dipolar excitons, such as MoSe$_2$/WSe$_2$ \cite{tran2019evidence,kunstmann2018momentumspace,alexeev2019resonantly,forg2021moire,wu2018theory} or WSe$_2$/WS$_2$ \cite{xiong2023correlated,park2023dipole,gao2024excitonic,lian2024valley,wang2023intercell}, as shown in Fig.\ref{fig: Fig 1} (a). 
Dipolar excitons in these heterobilayers can serve as charge sensors \cite{li2021local,miao2021strong,gao2024excitonic}, naturally suggesting the possibility to utilize them as probes of fractional charge. Going beyond the 
effective tight-binding model analysis of Ref.~\cite{lu2026exciton}, 
we treat exciton--quasihole binding by combining a mobile interlayer exciton interacting with electrons described by a material-specific continuum model of t-MoTe$_2$. This allows us to identify parameter regimes in which the spatial manipulation of anyons is within experimental reach, and reveals that the charge distribution in the anyon core, shaped by the moiré structure, deviates strongly from that in Landau-level systems.

Our analysis centers on a Fermi--Bose mixture in which a single exciton interacts with spin-polarized electrons in a partially filled moir\'e Chern band hosting a single quasihole of fractional charge $e/3$. Using exact diagonalization (ED), we determine how the binding energy and the exciton--quasihole entanglement depend on the exciton-electron interactions and the exciton dispersion. We uncover ground states formed by an entangled exciton--quasihole configuration in which the two constituents strongly overlap in real space, and we resolve moir\'e-induced structure of the quasihole charge distribution spread over the scale of the moir\'e unit cell, providing new insights into the nature of fractionalized excitations in moir\'e materials.
The main findings of this work are:
\\
\noindent (i) \textit{Binding energy and exciton--quasihole entanglement versus distance $h$ and exciton mobility.} Although the exciton--electron interaction is repulsive, the electronic charge depletion associated with a quasihole produces an effective attraction, signaled by a redshift of the exciton energy. We compute this redshift as a function of the distance $h$ between the FCI and the TMD heterobilayer and of the exciton bandwidth, which we tune through the exciton effective mass $m^{*}_{b}$. For small $m^{*}_{b}$, the exciton kinetic energy prevents binding: both the binding energy and the exciton--quasihole entanglement vanish. For large $m^{*}_{b}$, where the exciton bandwidth is small compared with the FCI gap, the exciton and quasihole are only weakly bound for $h \gtrsim 5$~nm. Binding strengthens markedly at shorter distances, reaching binding energies of order $1$~meV for $h \sim 1$--$2$~nm. This enhanced binding is accompanied by strong exciton--quasihole entanglement, with the entropy approaching the maximally allowed upper bound.
These trends are consistent with the recent observation of anyon--trions in t-MoTe$_2$ \cite{li2026signatures}, where trions confined in potential traps display photoluminescence features that are redshifted within slightly doped FCI states. In that setting, quantum confinement suppresses the kinetic energy of the excitonic complex, playing a role analogous to the large-$m^{*}_{b}$ regime identified here, while the vanishing layer separation corresponds to the strong-binding limit $h \to 0$.
\\

\noindent (ii) \textit{Spatial structure of the anyon core and quantization of the anyon charge.} From the exciton--electron (boson--fermion) correlation function, we determine the electronic charge distribution induced around the exciton. This distribution is sharply peaked near the exciton for small $h$ and large $m^{*}_{b}$, and it fades as the exciton--quasihole entanglement weakens and the binding energy vanishes. It therefore provides a direct real-space image of the binding process. The integrated charge depletion converges to the fractional quasihole charge $e/3$, confirming that the exciton pins the anyon in space.
The anyon charge distribution extends over a region comparable to the moir\'e unit cell and has a $C_3$-symmetric profile. Writing the charge depletion near the exciton as $D(\bs{r}) \approx \rho_{0}(r) + \rho_{3}(r)\cos(3\theta)$, the monopole component $\rho_{0}(r)$ carries the anyon charge, while $\rho_{3}(r)\cos(3\theta)$ is a higher-order multipole moment induced by the moir\'e potential. Remarkably, the charge distribution in the anyon core and the binding energy are sensitive on the quantum geometry of the bands, features that have no counterpart in the rotationally symmetric quasihole of the Landau-level FQH problem \cite{laughlin1983anomalous,feldman2021fractional}. 
Such rich structure in the quasihole core could be resolved by scanning tunneling microscopy \cite{park2026local,deng2026real}, opening new venues for fractional charge probing and manipulation.

\vspace{5 pt}
\noindent
\textit{Fermi-Bose Continuum Model --}
We consider spin/valley-polarized fermions described by a continuum model for moiré TMDs \cite{fengcheng2019topological}
\begin{equation}
h^{\rm f} 
= \left(\begin{array}{cc}
\frac{-\hbar^2\left(\boldsymbol{p}-\boldsymbol{k}_b\right)^2}{2 m^*}+V_b(\boldsymbol{r}) & T(\boldsymbol{r}) \\
T^{\dagger}(\boldsymbol{r}) & \frac{-\hbar^2\left(\boldsymbol{p}-\boldsymbol{k}_t\right)^2}{2 m^*}+V_t(\boldsymbol{r})
\end{array}\right)
\label{fermionic continuum model}
\end{equation}
with moir\'e intralayer potentials and interlayer tunneling terms given by
\begin{equation}
\begin{split}
V_b(\mathbf{r})&=V e^{i \psi}\left(e^{i \mathbf{g}_1 \cdot \mathbf{r}}+e^{i \mathbf{g}_3 \cdot \mathbf{r}}+e^{i \mathbf{g}_5 \cdot \mathbf{r}}\right)+\text { h.c. }\\
V_t(\mathbf{r}) &= V_b(-\mathbf{r})\\
T(\mathbf{r})&=w\left(1+e^{-i \mathbf{g}_2 \cdot \mathbf{r}}+e^{-i \mathbf{g}_3 \cdot \mathbf{r}}\right).
\end{split}
\label{moiré potentials}
\end{equation}
We use $V=20.8$ meV, $w=-23.8$ meV, $\psi=-107.7^\circ$, and $m_f^*=0.6m_e$~\cite{wang2024fractional}. The twist angle is $3.89^\circ$, giving moir\'e length scale $a_M\approx5.18$ nm and unit-cell area $A_M=\sqrt{3}a_M^2/2$.

To investigate how exciton mobility affects quasihole binding, we describe the interlayer exciton as a mobile boson in a moiré potential. We adopt $V_t(\bs{r})$ as a reference moiré potential and vary the exciton effective mass $m^*_b$ to control the exciton bandwidth,
\begin{equation}
\label{eq: exciton H}
    h^{\rm b}(\boldsymbol{p},\boldsymbol{r}) = \frac{\hbar^2\boldsymbol{p}^2}{2 m^*_b}+V_t(\boldsymbol{r}).
\end{equation}
The variation of $m^*_b$ tunes the strength of moiré confinement relative to the exciton kinetic energy, allowing us to explore the competition between exciton mobility and quasihole binding.

To stabilize a 1/3 FCI, we fill the topmost fermionic moiré Chern band with $\nu=1/3$ electrons (or $\nu=-2/3$ holes), then introduce a dual-gate screened Coulomb interaction for fermions, projected on the topmost moiré Chern band
\begin{equation}
    H^{\rm ff}=\frac{1}{2A}\sum_{\bs{q},\bs{g}} V^{\rm ff}(\boldsymbol{q}+\boldsymbol{g})
\colon\bar{\rho}^{f}(\bs{q}+\bs{g})
\bar{\rho}^{f}(-\bs{q}-\bs{g})\colon
\label{fermi-fermi interaction}
\end{equation}
where $\bs{q}$ is the momentum transfer in the moiré Brillouin zone and $\boldsymbol{g}$ is the momentum transfer in moiré reciprocal vectors, $V^{\rm ff}(\boldsymbol{p})=\frac{e^2}{2 \epsilon_0 \epsilon_r |\bs{p}|} \tanh(|\bs{p}|D/2)$, $A=N_1N_2 A_{\rm M}$ is the sample area, and $\bar{\rho}^{f}$ is the fermion density projected on the topmost moiré Chern band
$\bar{\rho}^{f}(\bs{q}+\bs{g}) = \sum_{\bs{k},\bs{G},\sigma}\Phi^{*\bs{G}+\bs{g}}_{\bs{k}+\bs{q},\sigma}
\Phi^{\bs{G}}_{\bs{k},\sigma}\,
f^{\dagger}_{\bs{k}+\bs{q}}f_{\bs{k}}$
where $f^{\dagger}_{\bs{k}}$ creates a band fermion at $\bs{k}$ in the topmost moiré Chern band, $\Phi^{\bs{G}}_{\bs{k},\sigma} = \langle\bs{k}+\bs{G},\sigma|\Phi\rangle$ is the corresponding Bloch wavefunction and $\sigma = t,b$ is the layer index.

\begin{figure}[!t]
    \centering
    \includegraphics[width =\linewidth]{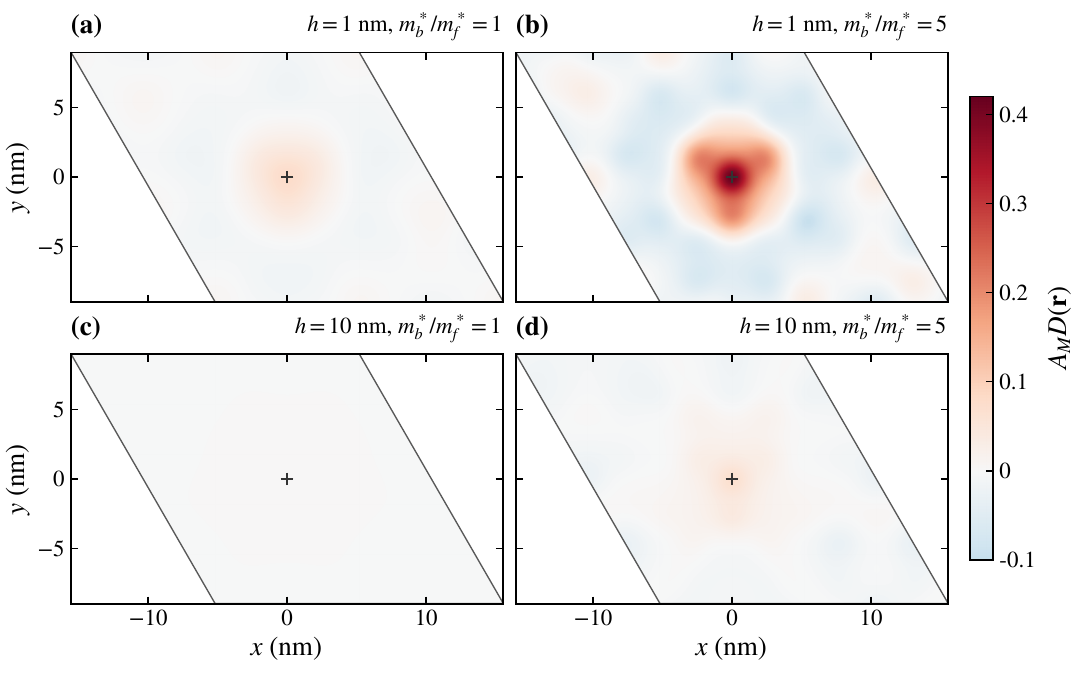}
\caption{Spatial structure of the exciton-induced electron depletion $D(\mathbf r)$ in the ground state, with $\mathbf r$ measured relative to the boson position. Panels show $h=1$ nm (top row) and $h=10$ nm (bottom row), for mass ratios $m_b^*/m_f^*=1$ (left column) and $5$ (right column). The shared color scale displays $A_M D(\mathbf r)$. Each parallelogram represents the $4\times4$ lattice containing five fermions and one $e/3$ quasihole. Crosses mark $\mathbf r=0$.}
\label{fig: Fig 2}
\end{figure}

\begin{figure}[!t]
    \centering
    \includegraphics[width = \linewidth]{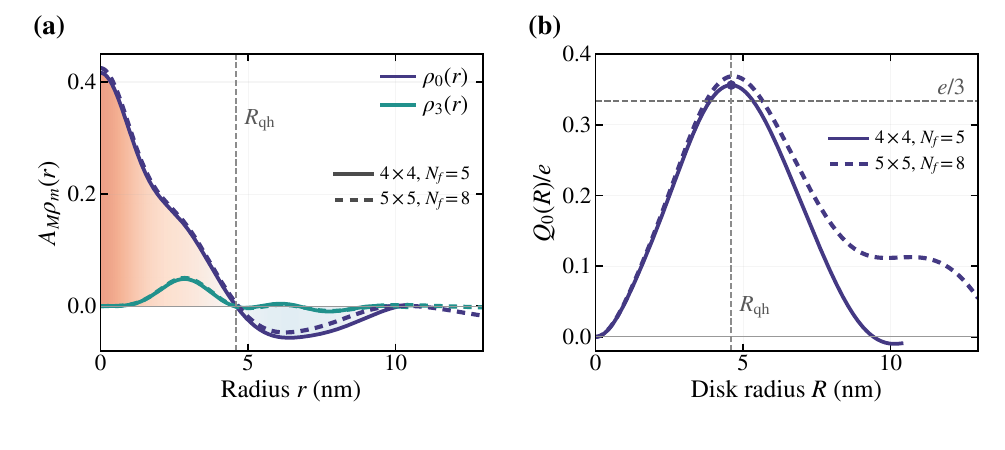}
\caption{Enclosed charge and angular structure of the depletion cloud. (a) Radial coefficients of the angular expansion. Here $\rho_0$ is the angular average, and $\rho_3$ describes the threefold, octupolar component. (b) Integrated depletion $Q_0(R)$ in a centered disk of radius $R$. Solid and dashed curves show the $4\times4$ and $5\times5$ systems, respectively. Shading highlights $\rho_0$ for the $4\times4$ system, and vertical dashed line denotes its anyon core radius $R_{\rm qh}$. Both panels use $h=1$ nm and $m_b^*/m_f^*=5$.}
\label{fig: Fig 3}
\end{figure}

\begin{figure}[!t]
    \centering
    \includegraphics[width = \linewidth]{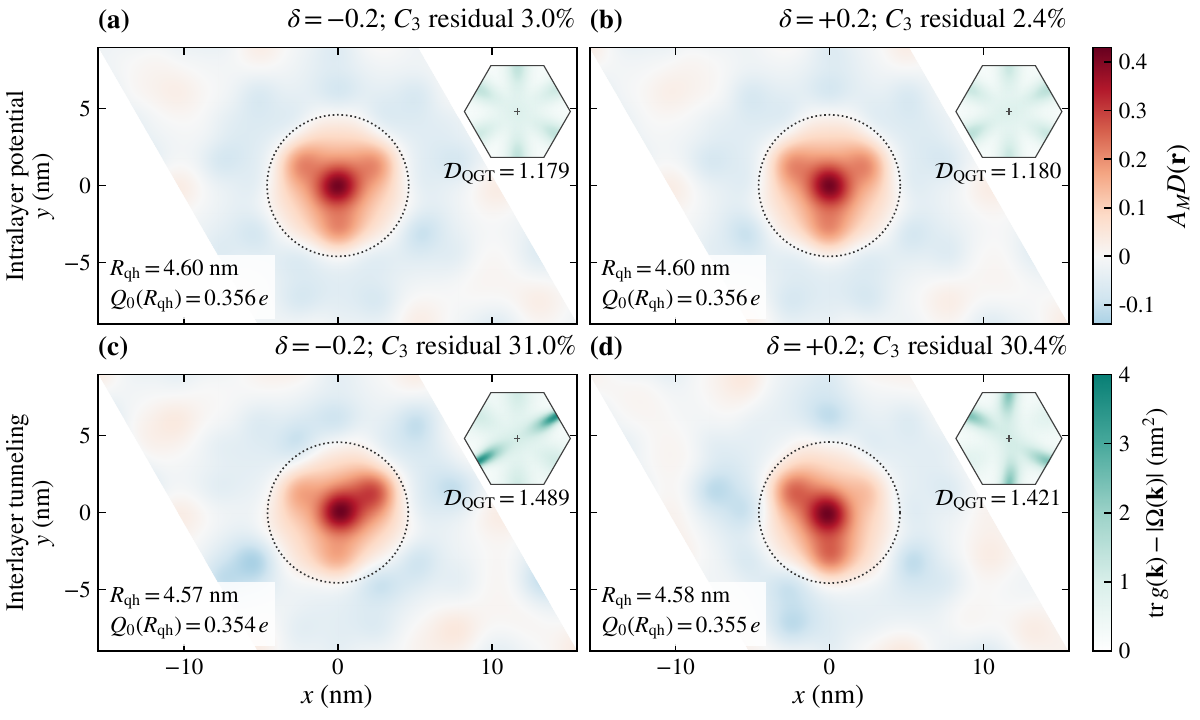}
\caption{Effect of moiré anisotropy on the quasihole depletion profile.  \(D(\mathbf r)\) for anisotropy in the fermion intralayer potential (a,b) and interlayer tunneling (c,d), with \(\delta=-0.2\) (left) and \(+0.2\) (right). Red and blue indicate electron depletion and excess, respectively. Dotted circles mark the quasihole core boundary \(R_{\mathrm{qh}}\), defined by the maximum of the disk-enclosed depletion \(Q_0(R)\). Each panel gives \(R_{\mathrm{qh}}\), \(Q_0(R_{\mathrm{qh}})\), and the \(C_3\)-symmetry residual within the core. Calculations use  \(h=1\,\mathrm{nm}\), \(m_b^*/m_f^*=5\). Insets plot the quantum geometry in first Brillouin zone.}
\label{fig: Fig 4}
\end{figure}

The exciton carries charges $-e$ and $+e$ at heights $h$ and $h+\ell$ above the fermion plane respectively, as shown in Fig.\,\ref{fig: Fig 1}(a). Such dipole orientation produces a repulsive electron--exciton interaction and an effective attraction to an electron deficit. Incorporating the relative permittivity $\varepsilon_{\rm r}$ and assuming a common polarizability $\alpha$ of the heterobilayer TMD hosting the exciton, we define $\lambda_0 = \frac{\alpha}{2\epsilon_0}$, and throughout Figs.\,\ref{fig: Fig 1}--\ref{fig: Fig 4}, we use $\epsilon_r=15$, $\lambda_0=5$ nm, $D=60$ nm, and $\ell=0.6$ nm. The dual-gate and sheet screened \cite{Marcel2025} dipolar fermi-bose interaction (see Supplemental Material\cite{SupplementalMaterial}) reads
\begin{equation}
    V^{\rm fb}(\boldsymbol{q})=\frac{V_{\rm dual\text{-}gate}(\boldsymbol{q})}
{1+(\lambda_0/\epsilon_r) \mathcal{K}_{\rm sheet}(q)}
\label{sheet screened dipole}
\end{equation}
where the bare dual-gate screened potential is
\begin{equation}
\begin{split}
V_{\rm dual\text{-}gate}(\boldsymbol{q}) = \frac{e^2}{2\epsilon_0\epsilon_r q}\left[
    (e^{-qh}-\frac{e^{qh}+e^{-qh}}{e^{qD}+1})\right.\\\left.
    -(e^{-q(h+\ell)}-\frac{e^{q(h+\ell)}+e^{-q(h+\ell)}}{e^{qD}+1})\right].
\end{split}
\end{equation}

We project the fermion-boson interaction to the topmost fermionic moiré Chern band and the lowest bosonic moir\'e band
{
\begin{equation}
H^{\rm fb}=\frac{1}{A}\sum_{\bs{q},\bs{g}} V^{\rm fb}(\boldsymbol{q}+\boldsymbol{g})
\colon\bar{\rho}^{f}(\bs{q}+\bs{g})
\bar{\rho}^{b}(-\bs{q}-\bs{g})\colon
\label{fermi-bose interaction}
\end{equation}
where
$\bar{\rho}^{b}(-\bs{q}-\bs{g}) = \sum_{\bs{k},\bs{G}}
\chi^{*\bs{G}-\bs{g}}_{\bs{k}-\bs{q}}
\chi^{\bs{G}}_{\bs{k}}
b^{\dagger}_{\bs{k}-\bs{q}}b_{\bs{k}}$
where $b^{\dagger}_{\bs{k}}$ creates a band boson at $\bs{k}$ in the lowest bosonic moiré band with dispersion $\varepsilon^{b}_{k}$, and $\chi^{\bs{G}}_{\bs{k}} = \langle\bs{k}+\bs{G}|\chi\rangle$ is the corresponding Bloch wavefunction.}
With that, the system Hamiltonian,
\begin{equation}
\label{eq: total H}
H = H^{\rm{ff}}+H^{\rm{fb}} + \sum_{\bs{k}}\varepsilon_{\bs{k}}b^{\dagger}_{\bs{k}}b_{\bs{k}}
\end{equation}
is diagonalized in the total center-of-mass momentum spanned by the basis \(|\mathcal{Q}\rangle = b^\dagger_{\boldsymbol{k}} \prod_{i=1}^{N_f} f^{\dagger}_{\boldsymbol{k}_i}|0\rangle\), where $\boldsymbol{k}+\sum_{i=1}^{N_f}\boldsymbol{k}_i\equiv\boldsymbol{Q}$ is the total momentum.
\vspace{5 pt}

\noindent
\textit{Spectral signatures --} 
In the background of the 1/3 FCI in a system with $N=N_1\times N_2$ moiré unit cells and $N_{\rm{e}} = N/3$ electrons, we excite a single $e/3$ quasihole by adding one unit cell (flux) to the system \cite{wu2012zoology,liu2015characterization,liu2013fractional,bernevig2012emergent}, resulting in a system with $N_{\rm qh}=N+1$ moiré unit cells. The binding energy between the exciton and the $e/3$ quasihole \cite{lu2026exciton, mostaan2025anyon}
\begin{equation}
    E_{\rm B} = (E_{\rm N}(V^{\rm fb})-E_{\rm N}(0))-(
    E_{\rm N_{qh}}(V^{\rm fb})-E_{\rm N_{qh}}(0))
    \,
\label{binding energy}
\end{equation}
is characterized by a red shift in the photoluminescence spectrum of the exciton.

Fig.~\ref{fig: Fig 1}(b) shows $E_{\rm B}$ as a function of the height $h$, after subtracting a uniform background that vanishes in the thermodynamic limit (see Supplemental Material\cite{SupplementalMaterial}).
{At larger separations $h\sim 10{\,\rm nm}$, the effective attraction weakens and the binding energy decreases toward zero. Binding is also suppressed for lighter exciton effective masses $m^*_b$, because broader exciton bandwidth increases the kinetic-energy cost of localization near the quasihole.}
Binding becomes stronger as two subsystems approach one another and as the boson-to-fermion effective mass ratio increases, reaching approximately $1.23$ meV at $h=1$ nm and $m_b^*/m_f^*=5$. At this mass ratio, the bandwidth of the lowest exciton band ($\approx 0.26$ meV) is substantially smaller than the FCI gap  ($\approx 2.72$ meV), a regime of strongly reduced exciton mobility that favors binding.

To quantify the exciton-quasihole correlations, we characterize the fermion-boson entanglement entropy $S_{\mathrm{fb}}=-\operatorname{Tr}\!\left(\rho_b\ln\rho_b\right)$ where $\rho_b=\operatorname{Tr}_{f}|\Psi\rangle\langle\Psi|$ is the reduced exciton density matrix and $|\Psi\rangle$ denotes the many-body ground state.
As shown in Fig.\,\ref{fig: Fig 1}(c), the entanglement entropy follows the same broad trend as the binding energy, where smaller height $h$ and larger exciton mass favors stronger entanglement. At $h=1$ nm and and $m_b^*/m^* = 5$, the entanglement saturates to $S_{\rm fb}\approx 2.45$, close to the maximum allowed $S_{\rm max}=\ln N_{\textrm{qh}}=\ln 16$, and furthermore, the ground state quasihole-manifold weight remains approximately \(97.7\%\)  
showing that strong entanglement can develop predominantly within the low-energy quasihole sector. Together, the binding energy and entanglement entropy support the formation of a correlated anyon bound state.
\\

\noindent
\textit{Spatial profile --}
Beyond these spectral and entanglement signatures, exciton--anyon binding
should also leave a clear imprint in real space. If the exciton binds a
quasihole, the electron density in its vicinity should be locally depleted
relative to the case in which the exciton and the electrons are uncorrelated.
To quantify this, we measure the electron density as seen from the exciton.

For an exciton located at $\mathbf R$, the conditional fermion density in the
ground state $|\Psi\rangle$ is
\begin{equation}
n_{f|b}(\mathbf r;\mathbf R)=
\frac{\langle\hat n_b(\mathbf R)\,\hat n_f(\mathbf R+\mathbf r)\rangle}
     {\langle\hat n_b(\mathbf R)\rangle},
\label{eq:conditional_density}
\end{equation}
where $\langle\cdots\rangle\equiv\langle\Psi|\cdots|\Psi\rangle$ and
$\mathbf r$ is the position relative to the exciton. Since the exciton is not
pinned to a particular location, we average Eq.~\eqref{eq:conditional_density}
over all exciton positions, weighted by the probability
$\langle\hat n_b(\mathbf R)\rangle/N_b$ of finding the exciton at $\mathbf R$.
This yields the fermion density profile in the exciton's frame,
\begin{equation}
n_{f|b}(\mathbf r)=
\frac{1}{N_b}\int_A d^2R\,
\langle\hat n_b(\mathbf R)\,\hat n_f(\mathbf R+\mathbf r)\rangle,
\label{eq:correlated_density}
\end{equation}
normalized such that $\int_A d^2r\, n_{f|b}(\mathbf r)=N_f$.

To isolate the part of this profile due to exciton--electron correlations, we
compare it with the profile obtained if the two densities were statistically
independent, i.e., replacing $\langle\hat n_b\hat n_f\rangle$ by
$\langle\hat n_b\rangle\langle\hat n_f\rangle$:
\begin{equation}
n_{\rm uncorr}(\mathbf r)=
\frac{1}{N_b}\int_A d^2R\,
\langle\hat n_b(\mathbf R)\rangle\,
\langle\hat n_f(\mathbf R+\mathbf r)\rangle .
\label{eq:uncorrelated_density}
\end{equation}
The difference
\begin{equation}
D(\mathbf r)=n_{\rm uncorr}(\mathbf r)-n_{f|b}(\mathbf r)
\label{eq:depletion}
\end{equation}
then measures the electron depletion induced by the exciton: $D(\mathbf r)>0$
near $\mathbf r=0$ signals a deficit of electrons surrounding the exciton,
as expected for a bound quasihole.
 
Fig.\,\ref{fig: Fig 2} shows the electron depletion Eq.~\eqref{eq:depletion} measured relative to the exciton position. At $h=1$ nm, increasing 
$m^*_b/m^*$ from 1 (Fig.\,\ref{fig: Fig 2}(a)) to 5 (Fig.\,\ref{fig: Fig 2}(b)) produces a pronounced depletion core centered beneath the exciton, consistent with the stronger binding energy and larger entanglement entropy in Fig.\,\ref{fig: Fig 1}. The quasihole core also develops an internal structure, revealing a threefold angular modulation associated with the moiré lattice. At $h=10$ nm, the depletion is strongly suppressed [Fig.\,\ref{fig: Fig 2}(c)-(d)]. These comparisons demonstrate how the competition between the fermi-bose interaction strength and exciton mobility controls the spatial correlations within the exciton-anyon composite, providing a remarkable real space resolution of the trapped anyon.

Because the number of fermions $N_f$ is fixed, the integral of $D(\bs{r})$ over the entire torus vanishes, for the positive core is accompanied by compensating electron excess at larger distances. Consequently, the fractional charge must be assessed by integrating over a region enclosing the core, rather than over the full system. The pronounced core in Fig.\,\ref{fig: Fig 2}(b) provides the basis for the enclosed-depletion and angular-multipole analyses presented in Fig.\,\ref{fig: Fig 3}.

The depletion of charge can be expanded into the monopole density and higher order multipole density as
\begin{equation}
\begin{split}
&D(r,\theta)=\rho_0(r)+\rho_3(r)\cos(3\theta)+\cdots\\
&\rho_0(r)=\frac{1}{2\pi}\int_0^{2\pi}D(r,\theta)\,d\theta,\\
&\rho_3(r)=\frac{1}{\pi}\int_0^{2\pi}D(r,\theta)\cos(3\theta)\,d\theta.
\end{split}
\label{multipole expansion}
\end{equation}
where the angular origin is chosen $30^\circ$ counterclockwise from the moiré lattice vector $\mathbf a_1$, for which the corresponding $\sin 3\theta$ component vanishes numerically in the profile shown, and the dipole and quadrupole density are negligible.

Fig.\,\ref{fig: Fig 3}(a) resolves the radial and angular structure of the quasihole depletion in Fig.\,\ref{fig: Fig 2}(b). The monopole density $\rho_0(r)$ peaks at the exciton position and decreases outward, becoming negative outside the quasihole core, whereas the octupolar component $\rho_3(r)$ vanishes at the origin and develops a maximum at a finite radius within the quasihole core. Thus, the internal structure visible in Fig.\,\ref{fig: Fig 2} reflects an angular modulation surrounding a predominantly isotropic central depletion. \textit{The angular dependence of the anyon core is a direct manifestation of the moiré lattice, and marks a departure from the rotation invariant quasihole profile in Landau levels \cite{laughlin1983anomalous}.} 
For comparison, Fig.~\ref{fig: Fig 3} includes $N_f=8$ fermions on a $5\times5$ lattice, again hosting a single $e/3$ quasihole. Within the core, $\rho_0$ and $\rho_3$ closely match the $4\times4$ results, showing that both the radial profile and the octupolar structure of the quasihole persist at the larger system size.

The depletion charge enclosed within a disk of radius $R$ is
\begin{equation}
Q_0(R)=e\int_{|\mathbf r|\le R}D(\mathbf r)\,d^2r
=2\pi e\int_0^R r\,\rho_0(r)\,dr ,
\end{equation}
and we define the quasihole core radius as the radius that maximizes it,
$R_{\rm qh}=\operatorname*{arg\,max}_R Q_0(R)$.
For the $4\times4$ system [Fig.~\ref{fig: Fig 3}(b)], $Q_0$ peaks at
$R_{\rm qh}\approx4.6$~nm with $Q_0(R_{\rm qh})\approx0.356\,e$, close to the
expected $e/3$. The $5\times5$ system gives nearly the same core,
$R_{\rm qh}\approx4.7$~nm and $Q_0(R_{\rm qh})\approx0.369\,e$.
The two sizes differ mainly beyond the core, where $Q_0(R)$ decays more slowly
in the larger system: the compensating electron excess spreads to larger
distances, while the quasihole core remains essentially unchanged. Increasing
the system size thus improves the separation between the localized depletion
and its compensating excess.
The maximum enclosed depletion also depends on the shape of the integration
window; a parallelogram window yields maxima closer to $e/3$ for both system
sizes (Fig.~\ref{figs2} in \cite{SupplementalMaterial}).

The threefold angular modulation of the core is quantified by the octupole
moment of the depletion within $R_{\rm qh}$,
\begin{equation}
\label{octupole moment}
M_3^{\rm qh}=e\int_{|\mathbf r|\le R_{\rm qh}}r^3e^{i3\theta}D(\mathbf r)\,d^2r
=\pi e\int_0^{R_{\rm qh}}r^4\rho_3(r)\,dr .
\end{equation}

For the $4\times4$ system we find
$M_3^{\rm qh}=1.04\,e\,\mathrm{nm}^3=7.46\times10^{-3}\,e\,a_M^3$.
To determine whether this octupolar structure is inherited from the
$C_3$ symmetry of the moir\'e Hamiltonian, we introduce a controlled
$C_3$-breaking anisotropy in the fermion sector (Fig.~\ref{fig: Fig 4}) by
independently perturbing the first-harmonic amplitudes of either the
intralayer moir\'e potential or the interlayer moir\'e tunneling
\begin{equation}
\begin{split}
&V_b^\delta(\bs{r}) = V^\delta_t(-\mathbf{r})\rightarrow V e^{i \psi}\left[(1+\delta)e^{i \mathbf{g}_1 \cdot \mathbf{r}}+\right.\\&\hspace{35pt}(1-\delta/2)e^{i \mathbf{g}_3 \cdot \mathbf{r}}+\left.(1-\delta/2)e^{i \mathbf{g}_5 \cdot \mathbf{r}}\right]+\text { h.c. }\\
&T^\delta(\mathbf r)\rightarrow w[
(1+\delta)
+(1-\frac{\delta}{2})\left(e^{-i\mathbf g_2\cdot\mathbf r}
+e^{-i\mathbf g_3\cdot\mathbf r}\right)]
\,,
\end{split}
\label{moire anisotropy}
\end{equation}
keeping
the mean amplitude of the three harmonics fixed, and tracking how the depletion
profile and the enclosed charge $Q_0(R)$ respond. 

We quantify the deviation from the threefold symmetry using the normalized residual
\begin{equation}
\eta_{C_3}=
\left[
\frac{\displaystyle\int_{|\mathbf r|<R_{\mathrm{qh}}}
\left|D(\mathbf r)-D(\mathcal R_{2\pi/3}\mathbf r)\right|^2\,d^2r}
{\displaystyle\int_{|\mathbf r|<R_{\mathrm{qh}}}
|D(\mathbf r)|^2\,d^2r}
\right]^{1/2},
\end{equation}
where $R_{2\pi/3}$ denotes a 120$^\circ$ rotation. The core radius $R_{\rm qh}$ is determined individually for each profile in Fig.\,\ref{fig: Fig 4}(a-d). Intralayer moiré potential anisotropy produces only modest changes, as shown in Fig.\,\ref{fig: Fig 4}(a,b), with residuals of $3.0\%$ and $2.4\%$ for negative and positive $\delta$, respectively. These values remain close to the approximately $2.0\%$ baseline residual at $\delta=0$.
By contrast, interlayer tunneling anisotropy produces a substantially stronger response, as shown in Fig.\,\ref{fig: Fig 4}(c,d) at $\delta = \mp 0.2$, where the three lobes become visibly imbalanced, and the residual rises to approximately $30\%$ for both signs of $\delta$. Nevertheless, the anyon charge changes little: $Q_0(R_{\rm qh })$ remains between $0.354e$ and $0.356e$, while $R_{\rm qh}$ varies only from $4.57$ to $4.60$ nm. Thus, a substantial deformation of the core can occur without a comparable change in its charge. 

The contrasting responses are also reflected in the quantum geometry of the underlying moir\'e Chern band. The insets of Fig.~\ref{fig: Fig 4} show $\operatorname{tr}g(\bs{k})-|\Omega(\bs{k})|$, where $g(\bs{k})$ is the quantum metric \cite{provost1980riemannian} and $\Omega(\bs{k})$ is the Berry curvature. This nonnegative quantity measures the local departure from the ideal trace condition\cite{roy14band,ledwith2023vortexability,jackson2015geometric,cano2026ideal,li2026abelian}, i.e. $\operatorname{tr}g=|\Omega|$, with the integrated deviation characterized by $\mathcal{D}_{\rm QGT}=\int_{\mathrm{BZ}}d^2k\,(\operatorname{tr}g-|\Omega|)$. Starting from $\mathcal{D}_{\rm QGT}=1.167$ in the isotropic system where $\delta=0$, intralayer potential anisotropy produces only a small increase to $\mathcal{D}_{\rm QGT}\approx1.18$ for either sign of $\delta$. Interlayer tunneling anisotropy instead generates pronounced features in the trace deviation and increases $\mathcal{D}_{\rm QGT}$ to $1.489$ and $1.421$ for $\delta=-0.2$ and $+0.2$. The fermionic band retains $|C|=1$ in all cases. These results link the stronger deformation of the quasihole core under tunneling anisotropy to a larger departure from ideal band geometry, despite the unchanged Chern number and robust anyon charge.

\vspace{5pt}
\noindent
\textit{Discussion--}
In this work we characterized the interaction between an dipolar exciton in a TMD heterobilayer in proximity with a t-MoTe$_2$ fractional Chern insulator layer supporting a $e/3$ quasihole. Employing a microscopic analysis accounting for the electronic states of t-MoTe$_2$, we demonstrated the formation of exciton-anyon bound states when the two layers are a few nanometers apart and the exciton mobility is suppressed. This microscopic calculation uncovers characteristic binding energies $\sim 0.5-1$ meV, which are comparable to those experimentally seen in anyon-trion systems \cite{li2026signatures}.
Furthermore, the real-space resolution of the anyon charge distribution bound to the exciton reveals strong deviations from Landau level systems, including an octupole moment induced by the moiré potential.
Notably, recent scanning tunneling spectroscopy experiments in graphene quantum Hall systems have resolved local signatures of trapped fractional excitations, demonstrating experimental access to the spatial and energetic structure of anyon bound states \cite{park2026local,deng2026real}. 
In particular, the sensitivity of the tunneling spectrum to confinement anisotropy highlights the importance of the spatial organization of fractional charge \cite{park2026local}. 
Our results identify a corresponding opportunity in moiré FCIs: binding anyons to excitonic complexes and probing their internal properties via optical and scanning techniques within reach in experimentally accessible moiré platforms.
\\

\noindent
\textit{Note added} -  During completion of this work, we became aware of related studies on impurity-bound anyons \cite{cui2026anyonic,adlong2026anyonsimpuritiesfractionalcharge}.
\\

\noindent
\textit{Acknowledgments --} We are grateful to Haydn Adlong, Arthur Christianen,
Eugene Demler, Ataç Imamoglu, Weijie Li, Johannes Motruk,
Erich Mueller, Carlos Sa de Melo, Ajit Srivastava, Raman Sohal, and Glenn Wagner for stimulating discussions. This research was supported by the U.S. Department of Energy, Office of Science, Basic Energy Sciences, under Award DE-SC0023327. These results were first presented at the DOE Theoretical Condensed Matter Physics PI meeting on September 08, 2026 in Washington D.C., and we also thank the participants of this meeting for stimulating discussions.
OpenAI's ChatGPT was used to support the development of code for figure generation. All AI-assisted code was independently reviewed, verified, and validated by the authors to ensure accuracy.
\bibliography{bibliography}

\clearpage
\onecolumngrid
\begin{center}
    \textbf{\large Supplemental Material for: \\ Anyon bound states in MoTe$_2$ Fractional Chern Insulators}
\end{center}

% Reset and prepend "S" to counters
\setcounter{equation}{0}
\setcounter{figure}{0}
\setcounter{table}{0}
\setcounter{page}{1}
\makeatletter
\renewcommand{\theequation}{S\arabic{equation}}
\renewcommand{\thefigure}{S\arabic{figure}}
\renewcommand{\thetable}{S\arabic{table}}
\renewcommand{\thepage}{S\arabic{page}}
\makeatother

\section{Uniform interaction offset in binding energy}
The binding energy in Eq.~\eqref{binding energy} compares the interaction-induced energy shifts in two samples with the same particle numbers but different sample areas. Adding one moiré unit cell creates a quasihole while also reduces the mean electron density. Consequently, even an exciton that responds only to the spatially averaged electron density has a lower interaction energy in the larger sample. This uniform contribution is independent of exciton--quasihole correlations and must be separated from the energy associated with their binding. Here we derive the offset subtracted in Fig.~\ref{fig: Fig 1}(b).

The momentum transfer is $\bs{p}=\bs{q}+\bs{g}$, where $\bs{q}$ is in the moiré Brillouin zone and $\bs{g}$ is a moiré reciprocal lattice vector. The uniform term corresponds to  $\bs{q}+\bs{g}\rightarrow 0 $, which is equivalent to $\bs{q}=\bs{g}=0$. At zero momentum transfer, the projected density operators both become particle number operators 
\begin{equation}
\begin{split}
\bar{\rho}^{f}(0) &= \sum_{\bs{k},\bs{G},\sigma}\Phi^{*\bs{G}}_{\bs{k},\sigma}
\Phi^{\bs{G}}_{\bs{k},\sigma}\,
f^{\dagger}_{\bs{k}}f_{\bs{k}} = \sum_{\bs{k}}
f^{\dagger}_{\bs{k}}f_{\bs{k}} = \hat{N}_f\\
\bar{\rho}^{b}(0) &= \sum_{\bs{k},\bs{G}}
\chi^{*\bs{G}}_{\bs{k}}
\chi^{\bs{G}}_{\bs{k}}
b^{\dagger}_{\bs{k}}b_{\bs{k}} = \sum_{\bs{k}}
b^{\dagger}_{\bs{k}}b_{\bs{k}} = \hat{N}_b
\end{split}
\end{equation}

Within a sector of fixed fermion number $N^f$ and boson number $N^b$, the fermi-bose interaction Eq.~\eqref{fermi-bose interaction} has a state-independent component
\begin{equation}
H_0^{\rm fb} = \frac{1}{A} V^{\rm fb}(0)
\colon\bar{\rho}^{f}(0)
\bar{\rho}^{b}(0)\colon
=\frac{1}{A}V^{\rm fb}(0)\hat{N}_f\hat{N}_b
=\frac{1}{A}
V^{\rm fb}(0)\,N_f N_b,
\label{uniform q=0}
\end{equation}
which is a constant shifting every many-body eigenvalue by the same amount without changing the many-body eigenstate. Eq.~\eqref{uniform q=0} contributes to the binding energy \eqref{binding energy} by a finite offset, arising from the different sample areas,
\begin{equation}
    E_{\rm B}^{(0)} = \frac{1}{A_M}
V^{\rm fb}(0)\,N_f N_b(\frac{1}{N}-\frac{1}{N_{\rm qh}}).
\label{uniform finite size}
\end{equation}
Subtracting Eq.~\eqref{uniform finite size} from \eqref{binding energy} correctly gives $E_{\rm B}-E_{\rm B}^{(0)}\rightarrow0$ in the weak coupling limit $h\rightarrow+\infty$.
For the dipole orientation used in the main text, both constituent charges lie above the fermion plane, at $h$ and $h+\ell$. The zero-momentum limit is
\begin{equation}
V^{\rm fb}(0)=\lim_{p\to0}V^{\rm fb}(p)
=\frac{e^2\ell}{2\epsilon_0\epsilon_r}
\label{dipole zero limit}
\end{equation}
which is independent of the height $h$, but dependent on the exciton dipole moment $e\ell$.

The offset \eqref{uniform finite size} is positive when $N_{\rm qh}>N$, i.e., the larger sample has a smaller mean electron density and hence a smaller uniform repulsion. For a single $e/3$ quasihole, $N_{\rm qh}=N+1$. With $N_f=N/3$ and one boson, Eq.~\eqref{uniform finite size} becomes
\begin{equation}
E_{\rm B}^{(0)}=\frac{V^{\rm fb}(0)}{3A_M(N+1)}.
\label{e/3 offset}
\end{equation}
The offset therefore vanishes as $1/N$ for increasing system sizes. For the five-fermion samples used in Fig.~\ref{fig: Fig 1}, $N=15$ and $N_{\rm qh}=16$, giving
\begin{equation}
E_{\rm B}^{(0)}=\frac{V^{\rm fb}(0)}{48A_M}
\approx0.32\;\mathrm{meV},
\label{e/3 offset value}
\end{equation}
where $A_M\approx23.279\;\mathrm{nm}^2$, $\ell=0.6$ nm, and $\epsilon_r=15$.

\section{On the screening by the dielectric exciton layers}
In this section, we derive the exciton–electron interaction in the presence of dielectric screening by the exciton-hosting layers and electrostatic screening by metallic dual gates. Starting with two polarizable sheets in an unbounded dielectric medium, we then incorporate the dual-gate boundary conditions and identify the approximations that lead to the compact interaction used in the main text.

We first consider, in the absence of dual-gate screening, a pair of charges $Q_1 = Q$ and $Q_2 = -Q$ located, respectively, at $\bs{r}_1 = (\bs{\rho}=0,z=z_1 = h+\ell)$ and $\bs{r}_2 = (\bs{\rho}=0,z=z_2 = h)$, and compute the potential they create on the $z=0$ plane. Here $\bs{\rho}$ is a vector on the plane. The $Q_1$ and $Q_2$ charges sit on planes with polarizability $\alpha_1$ and $\alpha_2$, respectively. 

We are after solving the Maxwell's equation:
\begin{equation}
\label{eq: Gauss law}
\varepsilon_{0}\varepsilon_{r}\bs{\nabla}\cdot\bs{E} = \rho(\bs{r})
\,
\end{equation}
where $\rho(\bs{r}) = \rho_{\textrm{ext}}(\bs{r})+ \rho_{\textrm{ind,1}}(\bs{r}) + \rho_{\textrm{ind,2}}(\bs{r})$ is the sum of the external charge densities, $\rho_{\textrm{ext}}(\bs{r}) = Q\,\delta(\bs{\rho})\left[ \delta(z-z_1) - \delta(z-z_2) \right]$, and those induced on layers 1 and 2, which read
\begin{equation}
\rho_{\textrm{ind,i}}(\bs{r})
=
\left( -\bs{\nabla}_{\bs{\rho}}\cdot\bs{P}_{i} \right) \, \delta(z-z_i)
\,, 
\end{equation}
where $\bs{P}_i$ is the polarization of the i-th plane. For an isotropic medium $\bs{P}_i = \alpha_i \, \bs{E}_{i}$. Using $\bs{E} = -\bs{\nabla}\phi$, Gauss law \eqref{eq: Gauss law}  reads
\begin{equation}
\label{eq: Gauss law detailed}
-\varepsilon_{0}\varepsilon_{r} (\nabla^{2}_{\bs{\rho}}+\partial^{2}_{z})\,\phi(\bs{\rho},z)
=
Q\,\delta(\bs{\rho})\left[ \delta(z-z_1) - \delta(z-z_2) \right]
+
\alpha_{1}\,\nabla^{2}_{\bs{\rho}}\phi\, \delta(z-z_1)
+
\alpha_{2}\,\nabla^{2}_{\bs{\rho}}\phi\, \delta(z-z_2)
\end{equation}

We introduce Fourier transforms
\begin{equation}
\label{3D FFT}
\begin{split}
&\,
\phi(\bs{\rho},z) = \int \frac{d^{2}\bs{q}}{(2\pi^2)}\int \frac{d k}{2 \pi} 
\,
e^{i(\bs{q}\cdot\bs{\rho} + k\,z)}
\,
\tilde{\phi}(\bs{q},k)
\\
&\,
\delta(\bs{\rho})\delta(z) = 
\int \frac{d^{2}\bs{q}}{(2\pi^2)}\int \frac{d k}{2 \pi} 
\,
e^{i(\bs{q}\cdot\bs{\rho} + k\,z)}
\end{split}
\,.
\end{equation}

We also notice that potential at the plane $z = z_0$ can be expressed as
\begin{equation}
\label{2D FFT}
\begin{split}
\phi(\bs{\rho},z_0) &\,= \int \frac{d^{2}\bs{q}}{(2\pi)^2}\int \frac{d k}{2 \pi} 
\,
e^{i(\bs{q}\cdot\bs{\rho} + k\,z_0)}
\,
\tilde{\phi}(\bs{q},k)    
=
\int \frac{d^{2}\bs{q}}{(2\pi)^2}
e^{i\bs{q}\cdot\bs{\rho}}
\,
\int \frac{d k}{2 \pi}
e^{i k\,z_0}
\tilde{\phi}(\bs{q},k)
\\
&\,
\equiv
\int \frac{d^{2}\bs{q}}{(2\pi)^2}
e^{i\bs{q}\cdot\bs{\rho}}
\,\tilde{\phi}_{2D}(\bs{q},z_0)  
\,,
\end{split}    
\end{equation}
where, in the last expression, $\tilde{\phi}_{2D}(\bs{q},z_0)$ is the Fourier transform of the potential on the $z = z_0$ plane. 

Since $\nabla_{\bs\rho}^2$ becomes $-q^2$, Eq.~\eqref{eq: Gauss law detailed} reads
\begin{equation}
\epsilon(q^2-\partial_z^2)\tilde\phi_{2D}(\bs q,z)
=Q[\delta(z-z_1)-\delta(z-z_2)]
-\sum_{i=1}^{2}\alpha_iq^2\tilde\phi_{2D}(\bs q,z_i)\delta(z-z_i)=\rho_{2D}(\bs q,z').
\label{in-plane poisson}
\end{equation}

We introduce the Green function $G(q;z,z')$ as the response of the operator $q^2-\partial_z^2$ to a source localized at $z=z'$,
\begin{equation}
(q^2-\partial_z^2)G(q;z,z')=\delta(z-z').
\label{green definition}
\end{equation}

The Green function satisfies the same boundary conditions as the potential. Once $G(q;z,z')$ is known, the solution takes the form
\begin{equation}
\begin{split}
\tilde\phi_{2D}(\bs q,z)=\frac{1}{\epsilon}\int dz'\,G(q;z,z')\tilde\rho_{2D}(\bs q,z')
\end{split}
\label{green potential}
\end{equation}

With Eqs.~\eqref{3D FFT}--\eqref{2D FFT}, Eq. \eqref{eq: Gauss law detailed} leads to
\begin{equation}
\label{eq: FT main}
\tilde{\phi}(\bs{q},k)
=
\frac
{Q}
{\varepsilon_{0}\varepsilon_{r}(\bs{q}^2+k^2)}
\left(
e^{-i k z_1}
-
e^{-i k z_2}
\right)
-
\frac
{\alpha_1 e^{-i k z_1}}
{\varepsilon_{0}\varepsilon_{r}(\bs{q}^2+k^2)}\bs{q}^2
\tilde{\phi}_{2D}(\bs{q},z_1)
-
\frac
{\alpha_2 e^{-i k z_2}}
{\varepsilon_{0}\varepsilon_{r}(\bs{q}^2+k^2)}\bs{q}^2
\tilde{\phi}_{2D}(\bs{q},z_2)
\end{equation}

We now introduce two length scales $\lambda_1 = \frac{\alpha_1}{2\varepsilon_0\varepsilon_{r}}$ and $\lambda_2 = \frac{\alpha_2}{2\varepsilon_0\varepsilon_{r}}$. Then, multiplying Eq. \eqref{eq: FT main} by $e^{i k z_1}$ and integrating over $\int \frac{dk}{2\pi}$ produces the equation 
\footnote
{
We have used $\int^{\infty}_{-\infty} \frac{dk}{2\pi} \frac{1}{k^2+q^2} = \frac{1}{2q}$, and 
$\int^{\infty}_{-\infty} \frac{dk}{2\pi} \frac{e^{i k \ell}}{k^2+q^2} = \frac{e^{-q \ell}}{2q}$, for $z_1 - z_2 = \ell > 0$.
}
\begin{equation}
\label{eq: phi_1 phi_2 equation 1}
(1+ \lambda_1 q)\tilde{\phi}_{2D}(\bs{q},z_1)
=
\frac
{Q}
{2\varepsilon_{0}\varepsilon_{r}}
\frac{(1-e^{-q\ell})}{q}
-\lambda_2\,q\,e^{-q\ell}\,\tilde{\phi}_{2D}(\bs{q},z_2)
\,.
\end{equation}

Similarly, multiplying Eq. \eqref{eq: FT main} by $e^{i k z_2}$ and integrating over $\int \frac{dk}{2\pi}$ produces the equation
\begin{equation}
\label{eq: phi_1 phi_2 equation 2}
(1+ \lambda_2 q)\tilde{\phi}_{2D}(\bs{q},z_2)
=
-\frac
{Q}
{2\varepsilon_{0}\varepsilon_{r}}
\frac{(1-e^{-q\ell})}{q}
-\lambda_1\,q\,e^{-q\ell}\,\tilde{\phi}_{2D}(\bs{q},z_1)
\,.
\end{equation}

Eqs.~\eqref{eq: phi_1 phi_2 equation 1}\eqref{eq: phi_1 phi_2 equation 2} can be summarized by
Eq.~\eqref{green potential}. In the absence of the dual gate screening, we have
\begin{equation}
G(q;z,z')= \int \frac{k}{2\pi}\frac{e^{ik(z-z')}}{q^2+k^2}=\frac{e^{-q|z-z'|}}{2q},
\quad G(q;\pm \infty,z')=0.
\end{equation}

And therefore
\begin{equation}
\begin{split}
\tilde\phi_{2D}(\bs q,z_i)&=\frac{1}{\varepsilon_0\varepsilon_r}\int dz'\,G(q;z_i,z')\tilde\rho_{2D}(\bs q,z')\\
&=\frac{1}{\varepsilon_0\varepsilon_r}\int dz'\,
\frac{e^{-q|z_i-z'|}}{2q}
\left[
Q[\delta(z'-z_1)-\delta(z'-z_2)]
-\sum_{j=1}^{2}\alpha_j q^2\tilde\phi_{2D}(\bs q,z_j)\delta(z'-z_j)\right]\\
&=
\frac{Qe^{-q|z_i-z_1|}}{2\varepsilon_0\varepsilon_rq}-
\frac{Qe^{-q|z_i-z_2|}}{2\varepsilon_0\varepsilon_rq}
-\sum_{j=1}^{2}
\lambda_j q\,\tilde\phi_{2D}(\bs q,z_j){e^{-q|z_i-z_j|}}
\end{split}
\end{equation}

Introduce
\begin{equation}
\begin{split}
&\,
a_{1}(q) = 1 + \lambda_1\,q \,, \quad  a_{2}(q) = 1 + \lambda_2\,q  \,,  
\\
&\,
b_{1}(q) = \lambda_1\,q\,e^{-q\ell} \,, \quad  b_{2}(q) = \lambda_2\,q\, e^{-q\ell}
\,,
\\
&\,f(q) = \frac
{Q}
{2\varepsilon_{0}\varepsilon_{r}}
\frac{(1-e^{-q\ell})}{q}
\,.
\end{split}    
\end{equation}

Eqs. \eqref{eq: phi_1 phi_2 equation 1}
and \eqref{eq: phi_1 phi_2 equation 2} can be recast as

\begin{equation}
\begin{pmatrix}
a_{1}(q) & b_{2}(q)
\\
b_{1}(q) & a_{2}(q)
\end{pmatrix}    
\,
\begin{pmatrix}
\tilde{\phi}_{2D}(\bs{q},z_1)    
\\
\tilde{\phi}_{2D}(\bs{q},z_2)    
\end{pmatrix}
=
\begin{pmatrix}
f(q)
\\
-f(q)
\end{pmatrix}
\,.
\end{equation}
The solution reads
\begin{equation}
\begin{split}
&\,
\tilde{\phi}_{2D}(\bs{q},z_1) 
=
\frac{a_{2}(q)+b_{2}(q)}{a_{1}(q)\, a_{2}(q) - b_{1}(q)\, b_{2}(q)}
\,f(q)
\\
&\,
\tilde{\phi}_{2D}(\bs{q},z_2) 
=
-\frac{(a_{1}(q)+b_{1}(q))}{a_{1}(q)\, a_{2}(q) - b_{1}(q)\, b_{2}(q)}
\,f(q)
\end{split}    
\end{equation}

Since $\tilde{\phi}_{2D}(\bs{q},z_1)$ and $\tilde{\phi}_{2D}(\bs{q},z_2) $ are now known, we can calculate $\tilde{\phi}_{2D}(\bs{q},z=0)$ by
integrating Eq. \eqref{eq: FT main} over $\int \frac{dk}{2\pi}$. The result is
\begin{equation}
\tilde{\phi}_{2D}(\bs{q},z=0)
=
-e^{-q\,h}
\Big[
f(q) + \lambda_{1}\,q\, e^{-q \ell}\,\tilde{\phi}_{2D}(\bs{q},z_1)
+
\lambda_2 q\, \tilde{\phi}_{2D}(\bs{q},z_2)
\Big]
\,.
\end{equation}

Henceforth, we drop the ``2D" subscript. Consider the limit $\lambda_1 = \lambda_2 = 0$, in which case
\begin{equation}
\tilde{\phi}_{0}(q) = -e^{-q \,h }\, f(q) = \frac{Q}{2 \varepsilon_0\varepsilon_{r}}
\,
\frac{
\left(
e^{-q (h + \ell)} - e^{-q h}
\right)}
{q}
\,,
\end{equation}
which represents the potential created by a pair of $\pm Q$ charges without any screening effects. Notice that in the limit $q\ell \ll 1$, $\tilde{\phi}_{0}(q) \approx -\frac{p}{2 \varepsilon_0} e^{-q \, h}$, where $p = Q\ell$ is the dipole moment.

Moreover, consider the limit $\lambda_1 = \lambda_2 = \lambda$. In this case, $\tilde{\phi}_{2D}(\bs{q},z_1) = - \tilde{\phi}_{2D}(\bs{q},z_2) = \frac{f(q)}{a(q) - b(q)}$, which leads to the potential 
\begin{equation}
\tilde{\phi}_{2D}(\bs{q},z=0)
=
\frac{\tilde{\phi}_{0}(q)}
{1 + \lambda q\, (1 - e^{-q\ell})}
\,.
\end{equation}

In the presence of grounded metallic gates at $z=\pm D/2$, where $D$ is the gate-to-gate separation, the Green function is now denoted by $G_D(q;z,z')$ for the gated case
\begin{equation}
(q^2-\partial_z^2)G_D(q;z,z')=\delta(z-z'),
\quad G_D(q;\pm D/2,z')=0.
\label{eq:gated green}
\end{equation}
which gives
\begin{equation}
G_D(q;z,z')=
\frac{\sinh[q(z_<+D/2)]\sinh[q(D/2-z_>)]}{q\sinh(qD)},
\qquad z_<=\min(z,z'),\quad z_>=\max(z,z').
\label{gated green}
\end{equation}
Introduce $K_{ij}(q)=2qG_D(q;z_i,z_j)$, with $z_0=0$. We can also write
\begin{equation}
K(q;z,z')=e^{-q|z-z'|}
\frac{[1-e^{-2q(z_<+D/2)}][1-e^{-2q(D/2-z_>)}]}{1-e^{-2qD}}
\label{gated green K}
\end{equation}

We list all $K_{ij}$ here
\begin{equation}
\begin{aligned}
K_{11}&=
\frac{
[1-e^{-q(D+2h+2\ell)}]
[1-e^{-q(D-2h-2\ell)}]
}{1-e^{-2qD}},\quad
K_{22}=
\frac{
[1-e^{-q(D+2h)}]
[1-e^{-q(D-2h)}]
}{1-e^{-2qD}},
\\
K_{12}&=
e^{-q\ell}
\frac{
[1-e^{-q(D+2h)}]
[1-e^{-q(D-2h-2\ell)}]
}{1-e^{-2qD}}\\
K_{02}&=
e^{-qh}
\frac{1-e^{-q(D-2h)}}{1+e^{-qD}},\quad 
K_{01}=
e^{-q(h+\ell)}
\frac{1-e^{-q(D-2h-2\ell)}}{1+e^{-qD}}.
\end{aligned}
\end{equation}

Substituting the sheet charges from Eq.~\eqref{in-plane poisson} into Eq.~\eqref{green potential} gives
\begin{equation}
\tilde\phi_{2D}(\bs q,z_i)
=\frac{Q}{2\epsilon q}(K_{i1}-K_{i2})
-\lambda_1qK_{i1}\tilde\phi_{2D}(\bs q,z_1)
-\lambda_2qK_{i2}\tilde\phi_{2D}(\bs q,z_2).
\label{gated-self-consistency}
\end{equation}
Evaluating this expression at $z_1$ and $z_2$, we obtain the same form of coupled equations as before, with the gated coefficients
\begin{equation}
\begin{split}
a_1^D(q)&=1+\lambda_1qK_{11},\qquad
 a_2^D(q)=1+\lambda_2qK_{22},\\
b_1^D(q)&=\lambda_1qK_{12},\qquad
 b_2^D(q)=\lambda_2qK_{12},\\
f_1^D(q)&=\frac{Q}{2\epsilon q}(K_{11}-K_{12}),\qquad
 f_2^D(q)=\frac{Q}{2\epsilon q}(K_{12}-K_{22}).
\end{split}
\label{gated-coefficients}
\end{equation}
These equations can be recast as
\begin{equation}
\begin{pmatrix}a_1^D(q)&b_2^D(q)\\b_1^D(q)&a_2^D(q)\end{pmatrix}
\begin{pmatrix}\tilde\phi_{2D}(\bs q,z_1)\\\tilde\phi_{2D}(\bs q,z_2)\end{pmatrix}
=\begin{pmatrix}f_1^D(q)\\f_2^D(q)\end{pmatrix}.
\label{gated-response-matrix}
\end{equation}
The solution reads
\begin{equation}
\begin{split}
\tilde\phi_{2D}(\bs q,z_1)&=
\frac{a_2^D(q)f_1^D(q)-b_2^D(q)f_2^D(q)}
{a_1^D(q)a_2^D(q)-b_1^D(q)b_2^D(q)},\\
\tilde\phi_{2D}(\bs q,z_2)&=
\frac{a_1^D(q)f_2^D(q)-b_1^D(q)f_1^D(q)}
{a_1^D(q)a_2^D(q)-b_1^D(q)b_2^D(q)}.
\end{split}
\label{gated-response-solution}
\end{equation}
Since the potentials at $z_1$ and $z_2$ are now known, we can calculate the potential on the fermion plane by setting $i=0$ in Eq.~\eqref{gated-self-consistency}:
\begin{equation}
\tilde\phi_{2D}(\bs q,0)=\frac{Q}{2\epsilon q}(K_{01}-K_{02})
-\lambda_1qK_{01}\tilde\phi_{2D}(\bs q,z_1)
-\lambda_2qK_{02}\tilde\phi_{2D}(\bs q,z_2).
\label{gated-potential-at-zero}
\end{equation}
Consider the case $\lambda_1=\lambda_2=\lambda=\lambda_0/\epsilon_r$, and introduce $t=\lambda q$. Substituting Eq.~\eqref{gated-response-solution} into Eq.~\eqref{gated-potential-at-zero} gives
\begin{equation}
\tilde\phi_{2D}(\bs q,0)=\frac{Q}{2\epsilon q}
\frac{K_{01}[1+t(K_{22}+K_{12})]-K_{02}[1+t(K_{11}+K_{12})]}
{(1+tK_{11})(1+tK_{22})-t^2K_{12}^2}.
\label{gated-potential-equal-sheets}
\end{equation}
For $Q=e$, multiplying by the electron charge $-e$ yields
\begin{equation}
V_{\rm full}^{\rm fb}(q)=\frac{e^2}{2\epsilon q}
\frac{K_{02}[1+t(K_{11}+K_{12})]-K_{01}[1+t(K_{22}+K_{12})]}
{(1+tK_{11})(1+tK_{22})-t^2K_{12}^2}.
\label{full-gated-sheet}
\end{equation}

For a thin heterobilayer and a pair of distant gates, i.e., $\frac{D}{2}-h\gg\ell$, we can approximate $K_{11}\approx K_{22}$, so that Eq.~\eqref{full-gated-sheet} becomes
\begin{equation}
V^{\rm fb}(q)=\frac{e^2}{2\epsilon q}
\frac{(K_{02}-K_{01})[1+t(K_{22}+K_{12})]}
{(1+tK_{22})^2-t^2K_{12}^2}
=\frac{e^2}{2\epsilon q}
\frac{K_{02}-K_{01}}
{1+t(K_{22}-K_{12})}
= \frac{V_{\rm dual\text{-}gate}(\boldsymbol{q})}{1+t\mathcal{K}_{\rm sheet}(q)}
\end{equation}
where
\begin{equation}
\mathcal{K}_{\rm sheet}(q) = K_{22}-K_{12}=
\frac{
4\sinh[q(D/2+h)]\,
\cosh[q(D/2-h-\ell/2)]\,
\sinh(q\ell/2)
}{\sinh(qD)}
\end{equation}

\section{Density response of the 1/3 FCI to an exciton}

In this section, we examine how the FCI responds to the repulsive exciton--electron interaction by considering five fermions and one exciton on a $3\times5$ moiré lattice, at exact $1/3$ filling. The corresponding $4\times4$ system in the main text contains one additional unit cell and hosts a pre-existing $e/3$ quasihole. We retain the same band Hamiltonians and interactions, and evaluate $D(\mathbf r)$ from the ground state using Eq.~\eqref{eq:depletion}.

\begin{figure}[!htb]
    \centering
    \includegraphics[width=0.6\linewidth]{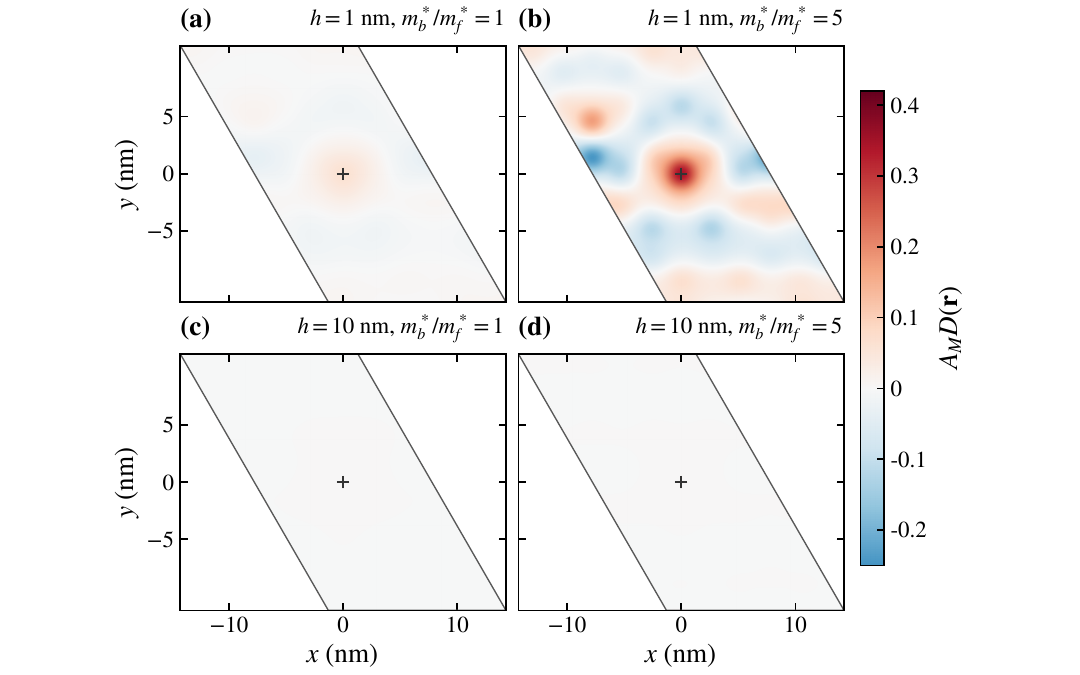}
    \caption{Exciton-induced electron depletion at exact $1/3$ filling, with five fermions on a $3\times5$ lattice. Panels show $h=1$ nm (top) and $10$ nm (bottom), for mass ratios 1 (left) and 5 (right). The color scale gives $A_M D(\mathbf r)$; positive and negative values denote electron depletion and excess, respectively. Crosses mark the exciton position.}
    \label{figs1}
\end{figure}

Although the isolated FCI has an excitation gap, the fermion--boson interaction can admix excited fermion states into the ground state, allowing the electron density to rearrange around the exciton. We quantify this admixture by the FCI-manifold weight $W_{\rm FCI}=\langle\Psi|(P_{\rm FCI}\otimes\mathbb{1}_b)|\Psi\rangle$, where $P_{\rm FCI}$ projects onto the three 1/3 FCI states of the isolated fermion Hamiltonian and $|\Psi\rangle$ is the ground state. At $h=1$ nm and $m_b^*/m_f^*=5$, $W_{\rm FCI}=0.827$ and the state remains predominantly within the FCI manifold, with a $17.3\%$ weight in excited fermion states. This admixture permits the repulsive interaction to reduce the electron density near the exciton while enhancing it elsewhere, producing the depletion and excess regions in Fig.~\ref{figs1}(b).

At $h=1$ nm, increasing $m_b^*/m_f^*$ from 1 to 5 lowers $W_{\rm FCI}$ from $0.989$ to $0.827$, while the central depletion grows from $A_M D(0)\approx0.063$ to $0.350$. A heavier exciton lowers the kinetic cost of developing spatial correlations with this density rearrangement. At $h=10$ nm, the weaker interaction leaves $W_{\rm FCI}>0.9999$ for both masses, and the depletion is correspondingly small, as shown in Fig.~\ref{figs1}(c,d).

We quantify the redistribution by integrating the charge depletion over a centered parallelogram,
\begin{equation}
Q_{\parallel}(L)=e\int_{\mathcal P(L)}d^2r\,D(\mathbf r).
\label{FCI enclosed charge}
\end{equation}
The window is a uniformly scaled copy of the corresponding torus, with edges parallel to the moiré lattice vectors. Here $L$ denotes the longer side and the $3\times5$ FCI window has sides $3L/5$ and $L$, while the $4\times4$ quasihole window has equal sides $L$. 

\begin{figure}[!htb]
    \centering
    \includegraphics[width=0.45\linewidth]{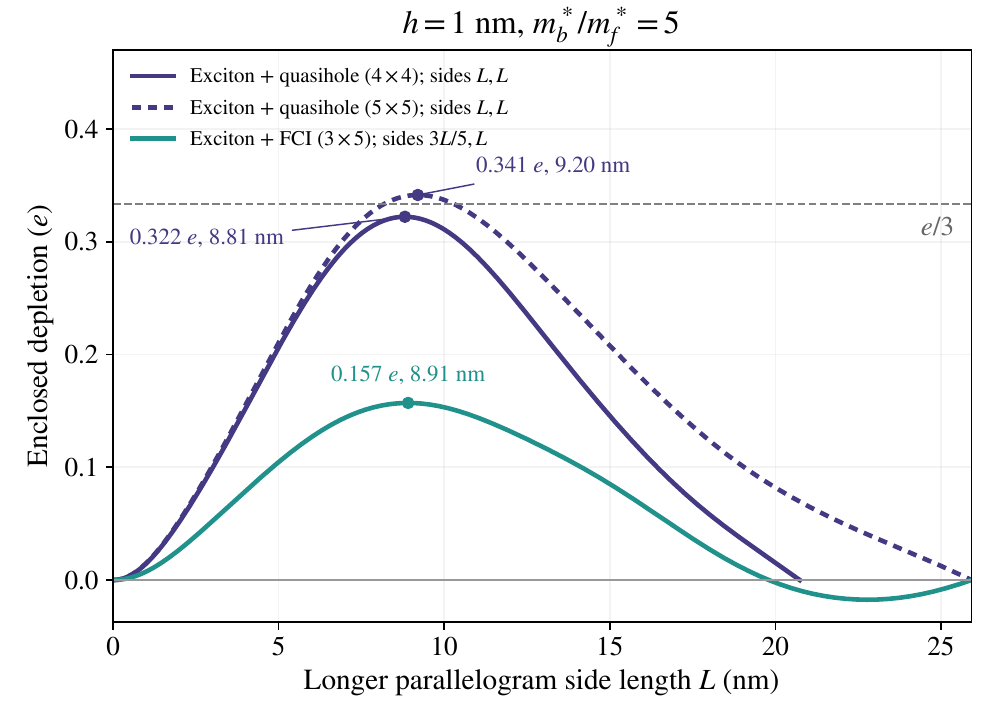}
    \caption{Enclosed depletion for the exciton--FCI ($3\times5$) and exciton--quasihole ($4\times4, 5\times5$) systems at $h=1$ nm and $m_b^*/m_f^*=5$, integrated over centered parallelogram windows. The horizontal axis is the longer side length $L$, and the legends specify both side lengths. Dots mark the maxima; the horizontal dashed line indicates $e/3$. Each curve ends when its window covers the full torus.}
    \label{figs2}
\end{figure}

At $h=1$ nm and $m_b^*/m_f^*=5$, the enclosed depletion reaches $0.157e$ at $L\approx8.91$ nm, as shown in  Fig.~\ref{figs2}, substantially below $e/3$. In comparison, the $4\times4$ system with a pre-existing quasihole reaches $0.322e$ at $L\approx8.81$ nm. Together with the dominant FCI-manifold weight, the smaller enclosed depletion is consistent with polarization of the FCI through an admixture of excited states, without a resolved $e/3$ quasihole core. Thus, the exciton repels the fermions and dresses itself with a local density redistribution, but the response has not developed the enclosed fractional charge characteristic of the bound quasihole compared to the main text.

For eight fermions on the $5\times5$ lattice, the parallelogram-enclosed depletion peaks at $0.341e$ for $L\approx9.20$ nm, closer to $e/3$ than the disk maximum of $0.369e$ in Fig.~\ref{fig: Fig 3}(b). The $4\times4$ parallelogram maximum, $0.322e$, is likewise closer to $e/3$ than its disk maximum of $0.356e$. This window dependence reflects the different portions of the anisotropic depletion and compensating electron excess included in the integral.

\end{document}